\documentclass[prl,showpacs,twocolumn]{revtex4}
\usepackage{graphicx}

\newcommand{\Teff}{\ensuremath{T_{\mathit{eff}}}}

\begin{document}

\title{Crumpling a Thin Sheet}
\author{Kittiwit Matan}
\author{Rachel Williams}
\author{Thomas A. Witten}
\author{Sidney R. Nagel}
\affiliation{The James Franck Institute and The Department of Physics, The
University of Chicago, Chicago, IL 60637}
\date{\today}

\begin{abstract}
Crumpled sheets have a surprisingly large resistance to further
compression.  We have studied the crumpling of thin sheets of Mylar under
different loading conditions.  When placed under a fixed compressive
force, the size of a crumpled material decreases logarithmically in time
for periods up to three weeks.  We also find hysteretic behavior when
measuring the compression as a function of  applied force.  By using a
pre-treating protocol, we control this hysteresis and find reproducible
scaling behavior for the size of the crumpled material as a function of
the applied force.
\end{abstract}

\pacs{68.60.Bs, 89.75.Da, 62.20.Fe, 46.32.+x}
\maketitle

If you, the reader, were to rip a page from this journal and crumple it,
squeezing it with your hands into a ball as hard as you can, the resulting
object is still more than $75\%$ air.  What gives this crumpled sheet its
surprising strength and how does the ultimate size of the sheet depend on
the forces applied?

The energy stored in crumpled sheets has been investigated in the limit
where the thickness, $\delta$, of the sheet is much less than the lateral
dimension,  $L$ (e.g., the diameter of a circular sheet)~\cite{1,2,3,4,5}.
Analysis of these objects has focused on two characteristic structures
caused by the crumpling: singular, conical points and the curved ridges,
which store most of the energy, connecting them.  As the sheet is crumpled
farther, the ridges must collapse so as to remain within the confining
container; and an increasing number of smaller ridges must form.  In the
limit where the diameter of the container, $D$, is much smaller than the
initial size of the sheet but still much larger than the sheet thickness
($\delta \ll D \ll L$) a scaling relation is predicted between the energy
stored and the size of the resulting crumple: $E \propto D^\beta$.  This
prediction assumes that the forces resisting crumpling are conservative.
One knows however, that frictional forces are involved as the sheet rubs
against itself and against the constraining walls.  How do these forces
affect the resulting energy balance?  In addition plastic flow can occur in
regions of high curvature effectively constraining the geometry of the
energy bearing structures.

The notion of scaling properties describing a crushed sheet goes back to
the initial experiments of Kantor et al.~\cite{5}  Further investigations
of such sheets were made by Gomes and collaborators~\cite{6,7}.  More
broadly, scaling behavior of force with compression in tenuous structures
was reported for random fibrous material by Baudequin et al.~\cite{8} and
for colloidal aggregates by several groups~\cite{9}.  Here we report
distinctive forms of scaling not seen in those previous studies.

To study how the external dimensions of a crumpled sheet depend on the
confining force, we have placed a large circular ($L = 34 cm$) sheet of
thin ($\delta = 12.5 \mu m$) aluminized Mylar~\cite{10} under a weighted
piston in a plastic cylindrical cell of diameter $D = 10.2 cm$.  The
cylinder is surrounded by a box to protect it from air currents in the room
and is mounted on an optical table to minimize the effects of vibration.
We measure compression by measuring the height, $h$, of the piston
above the base.  No special care is taken in the initial crumpling that
allows the large sheet to be placed within the cylinder.

\begin{figure}
\includegraphics[width=3.2in]{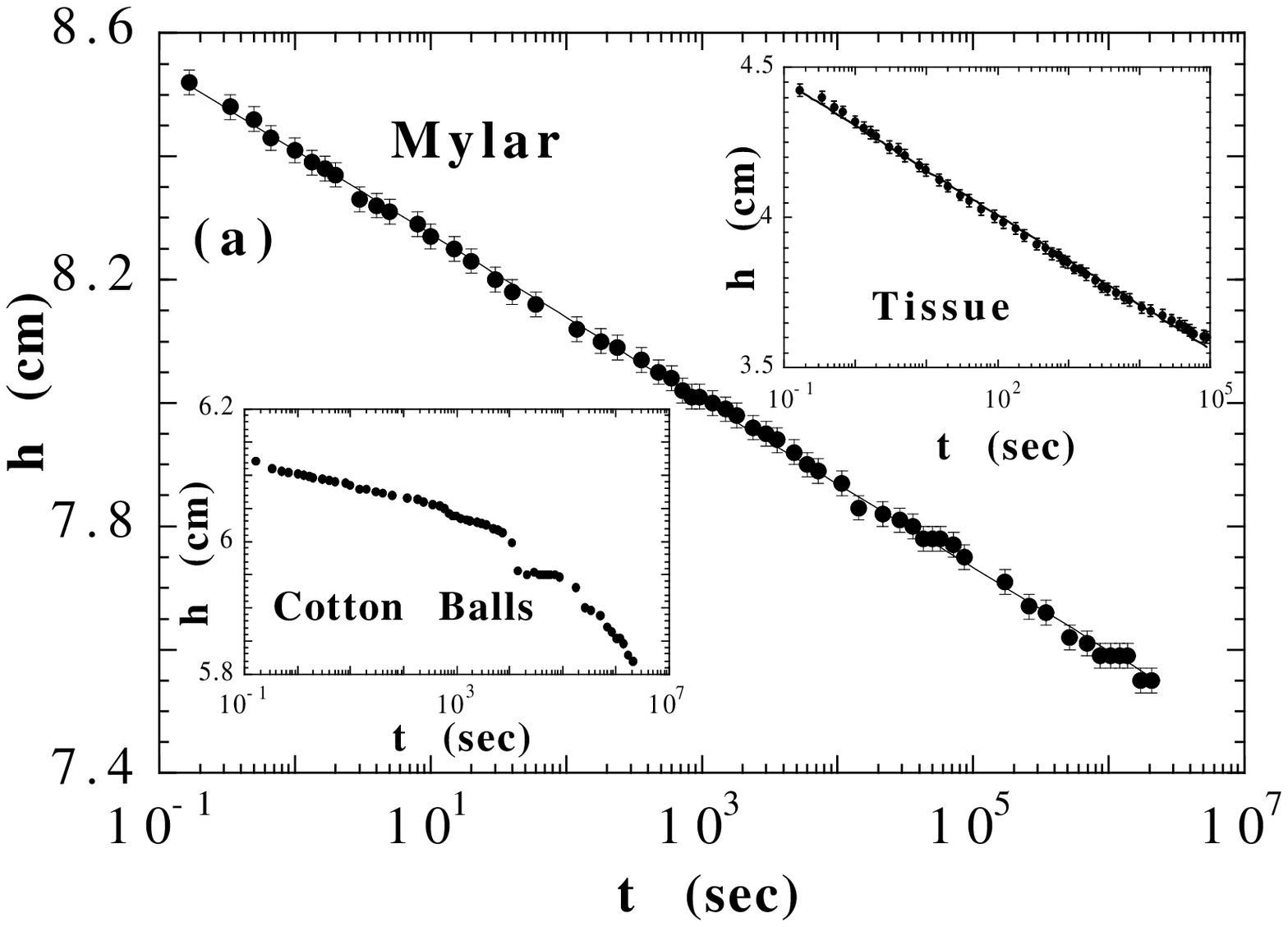}\\[10pt]
\includegraphics[width=3.2in]{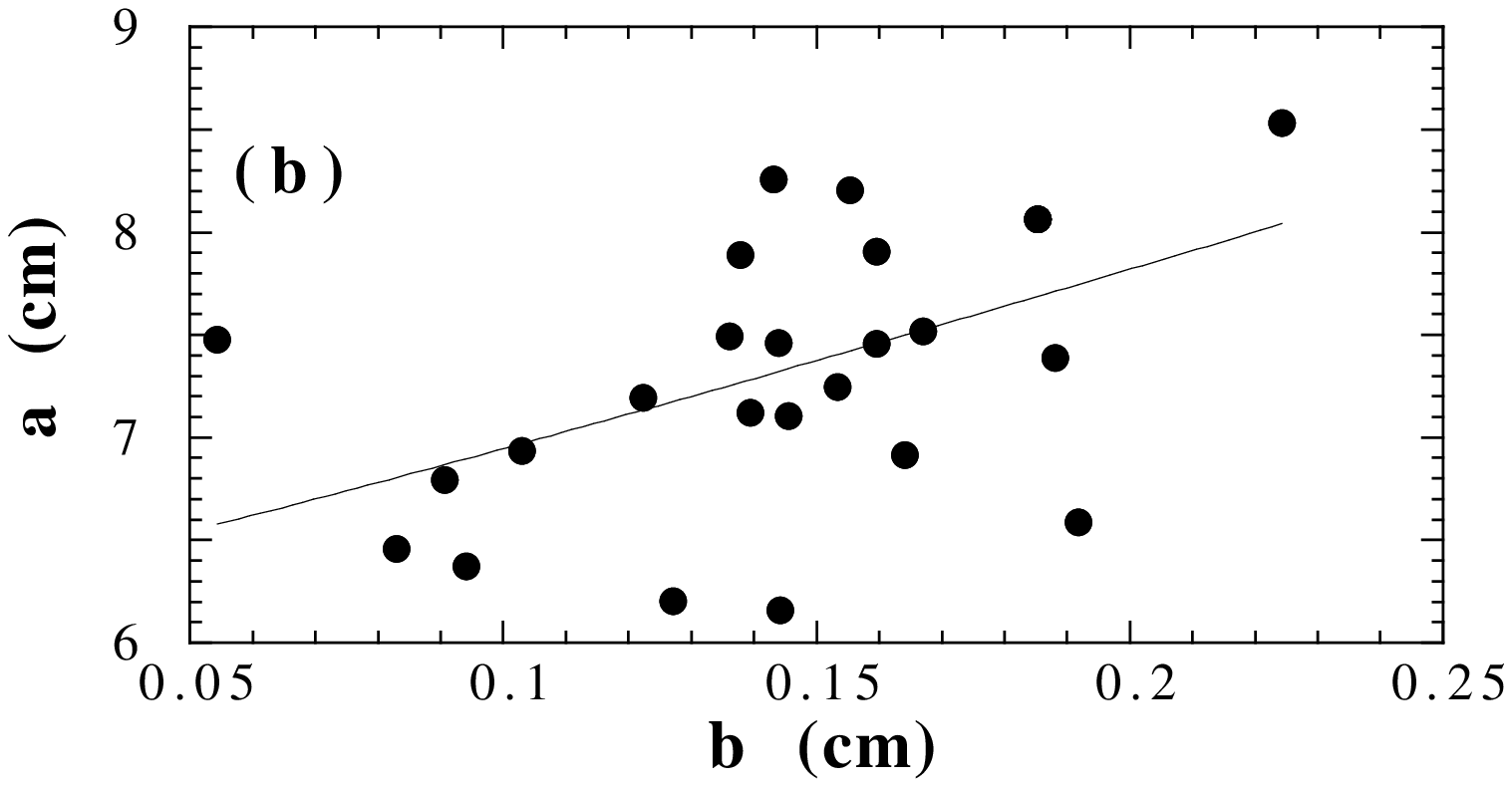}
\caption{(a)  Time-dependence of the height of a Mylar sheet compressed by
a mass $M = 200g$.  The straight line is a fit to Eq.~\ref{eqn1} with $a =
8.4 cm$ and $b = 0.14 cm$.  Insets show the time dependence of the height
for porous tissue and for cotton balls under the same conditions: $M =
200g$.  (b)  The constants $a$ and $b$ from Eq.~\ref{eqn1} for a series of
runs under nominally identical conditions.  No obvious evolution of the
values were found as the sheet was crumpled successively more times.
\label{fig1}
}
\end{figure}

After a mass, $M$, is placed on the piston, the piston settles to a new
height held up by the crumpled sheet below it.  However, this behavior is
not as simple as one might have expected: $h$ continues to decrease long
after the mass is initially introduced.  As shown in Fig.~\ref{fig1}(a), over
seven decades in time, $2 \times 10^{-1} sec < t < 2 \times 10^{6} sec$,
the entire duration of the experiment, the Mylar height decreases
logarithmically:
\begin{equation}
h = a - b \log (t/sec)
\label{eqn1}
\end{equation}
where $a$ and $b$ are constants.  Even after three weeks the piston did not
reach its asymptotic height.  Similar relaxation occurred in other
tenuous materials, as shown in the insets to Fig.~\ref{fig1}(a).  Absorbent
tissue paper~\cite{11} showed a $log t$ dependence like
that of the Mylar.  Cotton balls~\cite{12} showed a slightly smaller
relaxation over several weeks, in which the logarithmic dependence extended
only up to $10^4 sec$.   Although this logarithmic behavior is robust and
is found each time the Mylar is compressed, the constants $a$ and $b$
fluctuate between runs with no monotonic evolution of those
constants as the sheet is crumpled more times.  Fig.~\ref{fig1}(b) shows a
weak correlation between $a$ and $b$ for the same $200 g$ load on the
piston.

Albuquerque and Gomes~\cite{7} studied stress relaxation of crumpled
aluminum foil under fixed strain and found stretched-exponential behavior
quite different from the logarithmic dependence of strain that we measured
at fixed stress.  Aluminum  foil is much more maleable than the Mylar we
have used and presumably has more plastic flow along the ridges and
vertices.  This may be the origin of the different relaxation kinetics.
Menon~\cite{13} has found similar $\log t$ behavior as our results in stress
relaxation.

\begin{figure}
\includegraphics[width=3.2in]{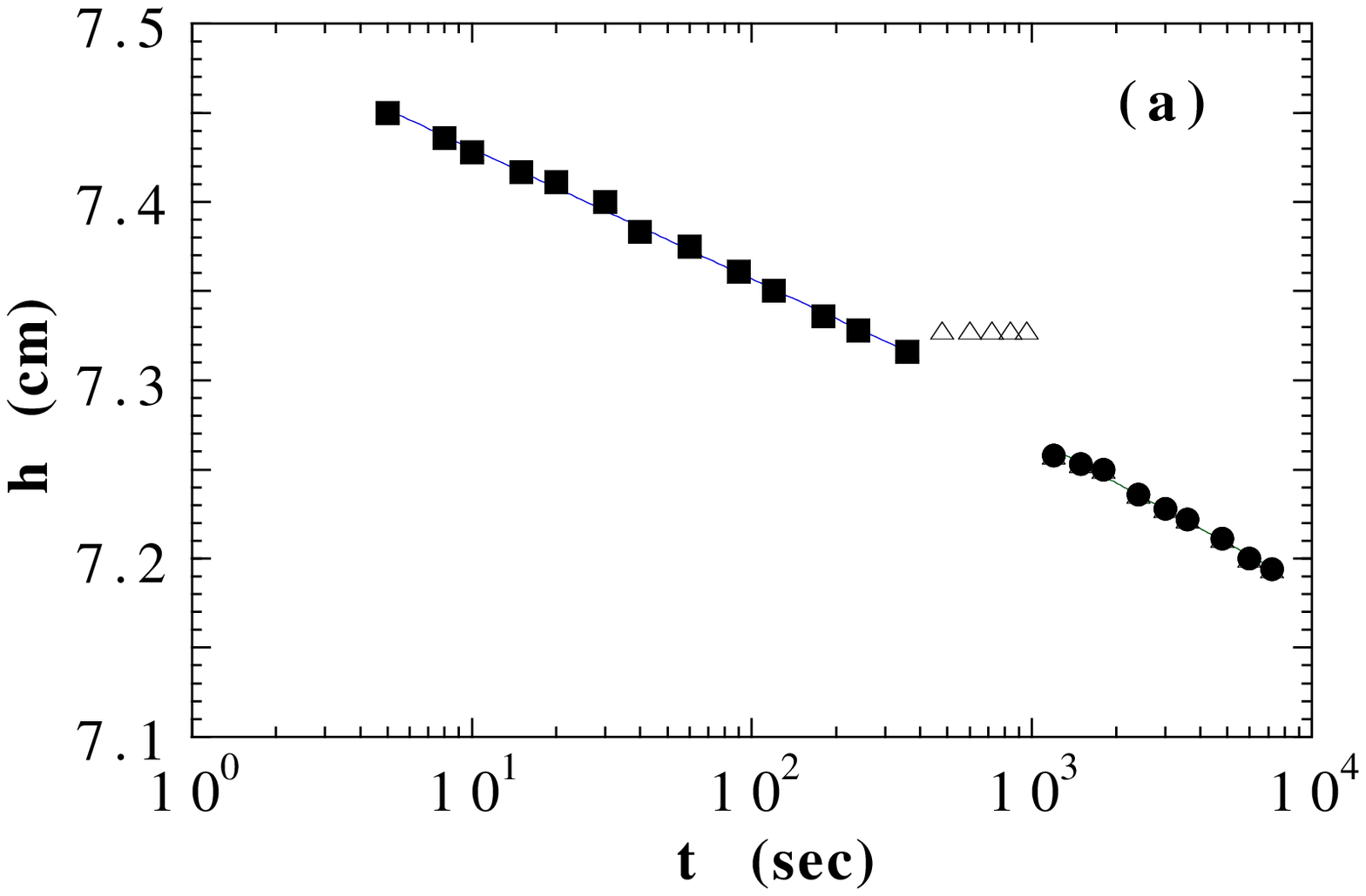}\\[10pt]
\includegraphics[width=3.2in]{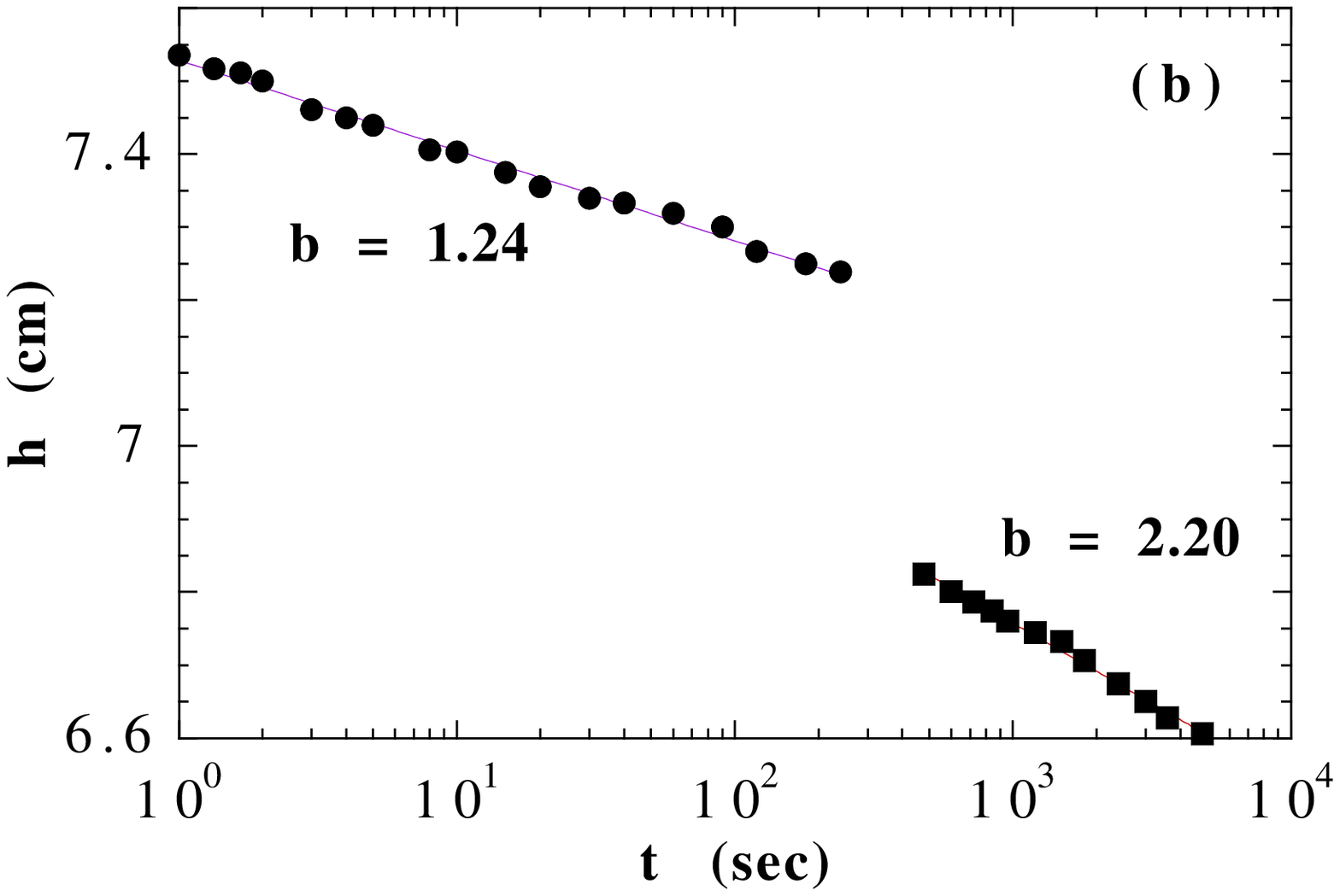}
\caption{(a)  Time-dependence of the height of a compressed Mylar sheet
before and after the compression was stopped.   The initial time-dependence
of the height of a Mylar sheet after a $200 g$ mass has been place on it is
shown by the filled squares.  After approximately $500 sec$. the piston is
lifted and fixed in place.  After $1000 sec$, the piston is released and
the subsequent relaxation is shown by the filled circles.  (b)  $h(t)$ for
a compressed Mylar sheet before and after external vibrations are
introduced.   Initially the container was at rest.  After approximately
$200 sec$, lateral vibrations with an acceleration of $2.5 m/sec^2$ at
$7.5 Hz$ were begun.  The resulting logarithmic slope, $b$, nearly doubled.
\label{fig2}
}
\end{figure}

Relaxation continues even if the compression is stopped.
Fig.~\ref{fig2}(a) shows the time dependence of the height of a Mylar sheet
after a $200 g$ mass has been place on it.  After $500 sec$. the piston is
lifted very slightly and fixed in place so that it can no longer compress
the material.  After $1000 sec$, the piston is released and quickly
sinks to a new position comparable to where it would have been had it never
been restrained; thereafter it relaxes at nearly the same
logarithmic rate as it did initially.

What is responsible for the slow decrease in height over these extended
periods?  External vibrations affect the compression.
Fig.~\ref{fig2}(b) shows $h(t)$ both before and after the container is
vibrated \textit{laterally} with an acceleration of $2.5 m/sec^{2}$ at  $7.5
Hz$.  Initially, the container was at rest and the logarithmic slope, $b$,
was determined.
After approximately $200 sec$, vibrations were begun and $b$ nearly
doubled.  This change, while substantial, is not so large as
to indicate that the slope in the quiescent state is caused by the ambient
vibrations in the room which are less than $10^{-2} m/sec^2$.

This logarithmic decay of  $h$ suggests activated relaxation (with an
effective temperature \Teff\ introduced to account for external
vibrations~\cite{14}) of the sheet between metastable minima separated by
energy barriers, $\Delta E$:  $dh/dt = -\zeta \exp [-\Delta E/(k_B
\Teff)]$.  As $h$ decreases, the crumpled sheet becomes progressively
stiffer suggesting that the barriers are increasing.  To lowest order
(i.e., $\Delta E \approx E_0 - C h$) we find:
\begin{equation}
dh/dt = -\zeta \exp [-(E_0 - C h)/(k_B \Teff)]
\label{eqn2}
\end{equation}
which has a solution of the form of Eq.~\ref{eqn1} where $a = d_1 + E_0/C
- (k_B \Teff/C) \ln(C \zeta /k_B \Teff)$ and $b = k_B \Teff /C$
  where $d_1$ is a constant.  As \Teff\ increases (e.g., due to
external vibrations) the logarithmic slope, $b$, also increases as
observed in the experiments.  If the precise initial crumpling determines
$E_0$ and $C$, then $a$ and $b$ should be correlated between runs: $a =
d_1 + b (\ln b + d_2)$ where $d_2 = E_0/k_B \Teff - \ln \zeta$.  The
curve in Fig.~\ref{fig1}(b) shows the fit with $d_1 = 6.2$ and $d_2 =
9.7$.   We note however, that $d_1$ and $d_2$ may also vary between
trials.

So far, we have reported the time dependence of the height, $h$, under a
fixed weight.  We are also interested in determining the effect of the
piston mass, $M$, on the height.  There are two problems which interfere
with obtaining such data in a straight-forward manner.  First, as we have
emphasized, $h$ varies logarithmically with time and it is thus impractical
to find the final height of a crumpled sheet under a given load.  Second,
there is large hysteresis so that different loading procedures produce
different values of $h$.  Such hysteresis is already suggested by the
variation between runs of  the constants, $a$ and $b$, describing the
relaxation.  Despite these problems we can obtain reproducible
results by following two protocols for handling the crumpled sheets.

In order to avoid the problems created by the slow relaxation, we
measure the height after a fixed interval of time (e.g., $100 sec$) and use
this value, instead of the unattainable asymptotic ($t = \infty$) value,
for comparison between different loadings.  Such a procedure
means that we neglect relaxation processes slower than $100 sec$.  To
obtain meaningful results we confine ourselves to comparing the piston
height for relatively large changes in the mass, $\Delta M$ so that $\Delta
h_{\Delta M}$ (the change in height measured at $t = 100 sec$ due to the
increased piston mass) is much greater than $\Delta h_{\Delta t}$ (the
change in height due to waiting extra time after the mass has
been added, e.g., between $100 sec$ and $1 day$ ):  $\Delta h_{\Delta M}
\gg \Delta h_{\Delta t}$.  In what follows we measure $h_{100}(M)$, the
height at $t = 100 sec$ after a mass, $M$, has been added to the piston.

Dealing with the second problem of hysteresis requires a protocol for
preparing a sheet prior to measurement.   Hysteresis is evident if we
start with a small initial mass and measure (as prescribed above) the
height after subsequent increments $\Delta M$ have been added until a large
final mass is applied.  If we then remove this load and repeat the
measurement with the initial mass, we find that the newly measured height
is smaller than that initially measured for the same mass.

\begin{figure}
\includegraphics[width=3.2in]{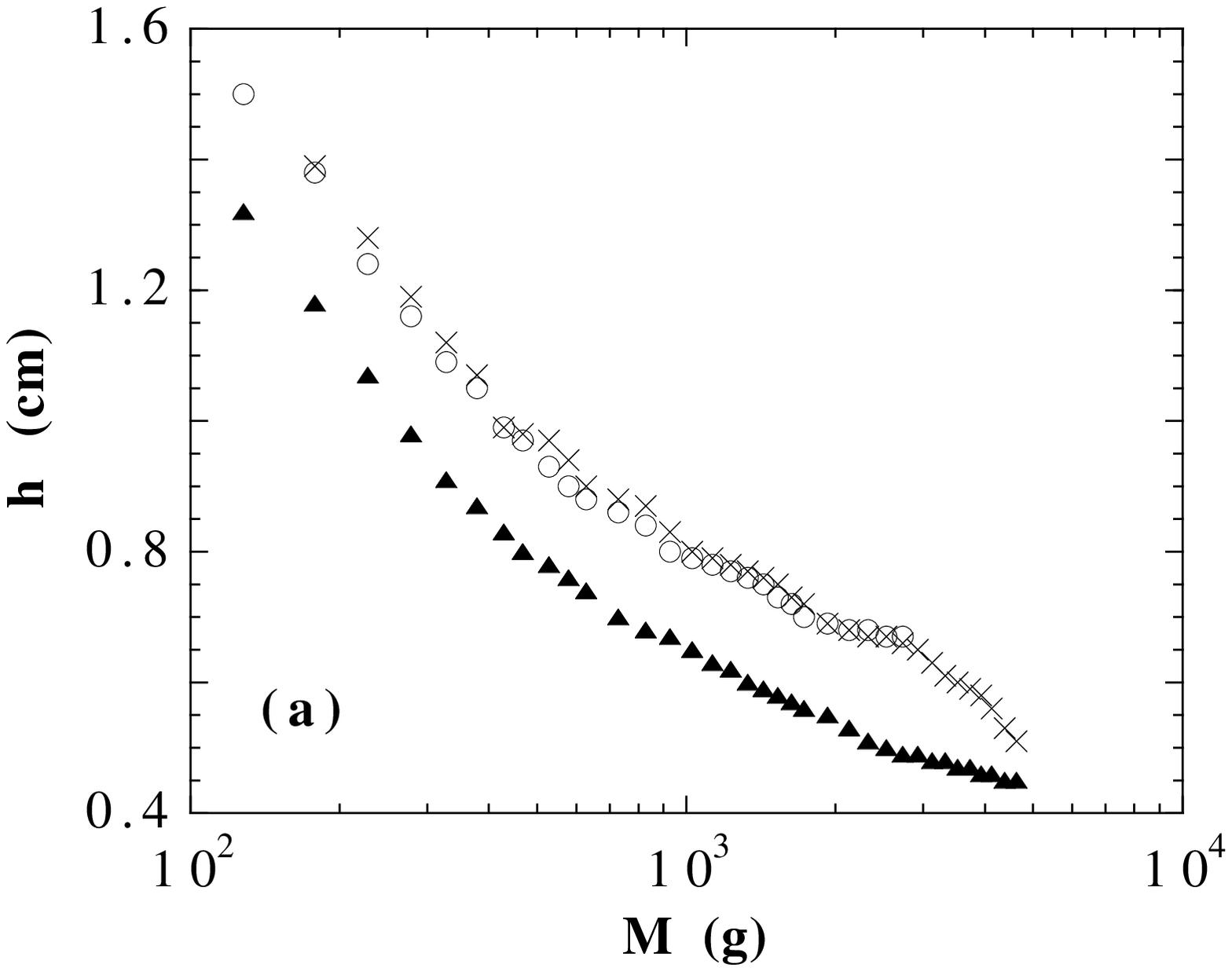}\\[10pt]
\includegraphics[width=3.2in]{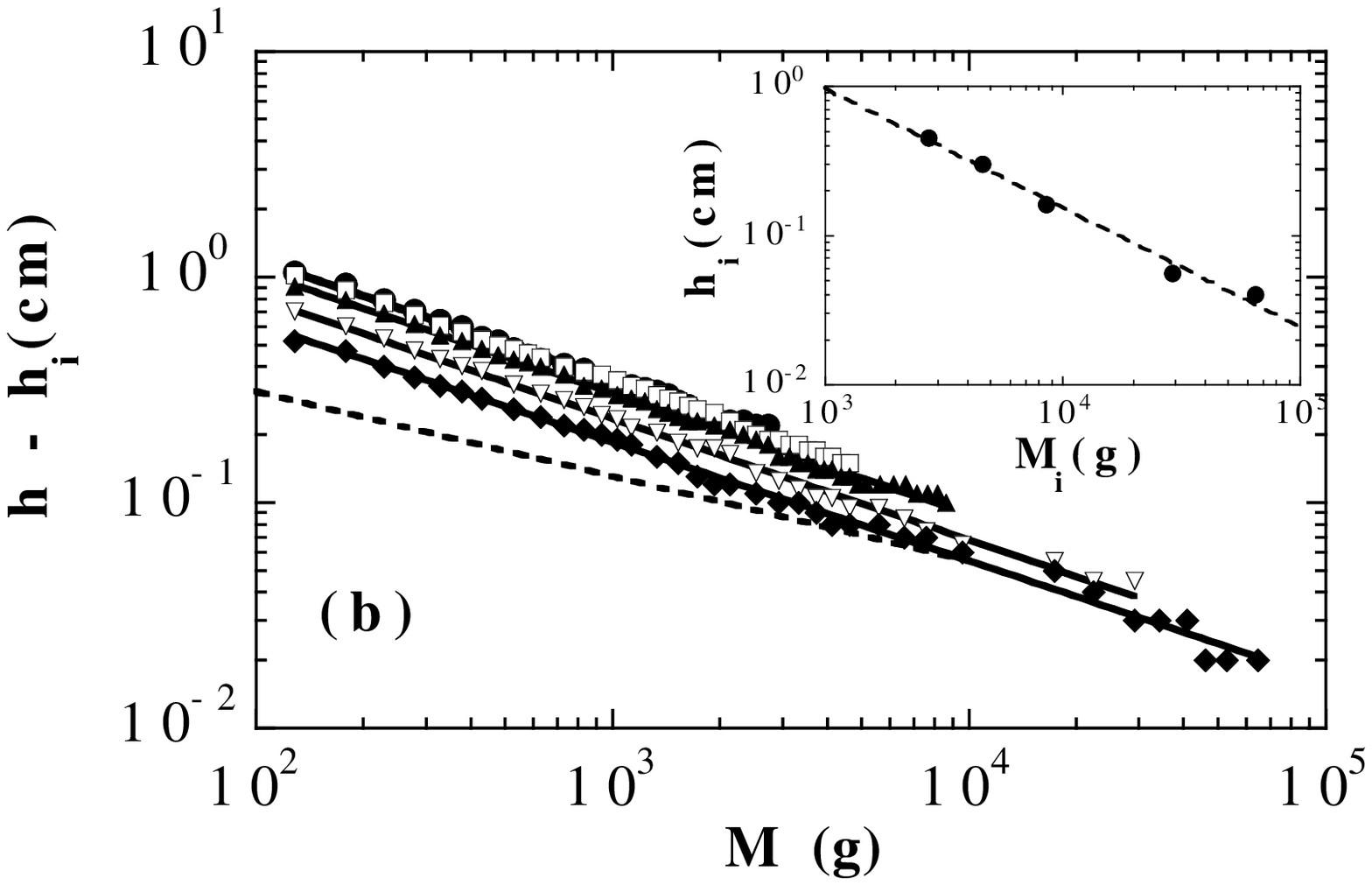}
\caption{(a)  Hysteresis in the height versus mass.  A crumpled Mylar sheet
is first placed under a mass $M_1 = 2.6 kg$ for one day.  The mass is then
removed and $h_{100}(M)$ is measured for $M < M_1$ (open O's).  If $M_1$ is
not exceeded, this curve is reproducible (X's).  If $M_1$ is exceeded, a
different curve $h_{100}(M)$ is found (solid triangles).  (b)  $\log
(h_{100}(M,M_i)-h_i)$ versus $\log M$ for five separate pre-treating masses
$M_i$.   Solid lines are fits to Eq.~\ref{eqn3} with $\alpha  = 0.53$.
Dashed line shows the predicted $h(t)$.  Inset shows scaling between the
asymptotic height, $h_i$, and the pre-treating mass, $M_i$.
\label{fig3}
}
\end{figure}

In order to avoid this hysteresis, we have followed the protocol of
Baudequin et al. for studying compressed glass wool~\cite{8}.  We pre-train
a sheet by first crumpling it under a mass $M_1$ for a period of one day.
The mass is then removed and we measure $h_{100}(M)$ for $M < M_1$ (as
shown for $M_1 = 2.6 kg$ by the open O's in Fig.~\ref{fig3}(a)).  This curve
is reproducible (shown by the X's) as long
as $M_1$ is not exceeded.  However, if we exceed $M_1$ and place a larger
mass on the piston, we no longer obtain the same curve.  With this larger
mass $M_2 = 4.5 kg$ we obtain a new curve (shown by the
solid triangles) that is itself reproducible as long as $M_2$ is not
exceeded.  We continue this measurement to obtain a family of curves
each of which corresponds to a pre-training mass $M_i$.  For each curve,
considered separately, we find a scaling relation:
\begin{equation}
h_{100}(M, M_i) = h_i + c M^{-\alpha}
\label{eqn3}
\end{equation}
where $h_i$ is the asymptotic value of the height for the pre-treating mass
$M_i$.  Such a family of curves is shown in Fig.~\ref{fig3}(b).  We find
$\alpha  = 0.53\pm{0.04}$ for all pre-treating masses.  This is smaller
than the value $\alpha  = 0.67$ found by Baudequin et al.~\cite{8} for
fibrous glass wool.  The inset to Fig.~\ref{fig3}(b) shows
a scaling relation between the asymptotic height, $h_i$, and $M_i$: $h_i
\propto M_i^{-\gamma}$ with $\gamma \approx
0.8$.  Despite the problems associated with the
logarithmic relaxation and hysteresis, Fig.~\ref{fig3}(b) shows that it is
possible, under well-defined
conditions, to obtain reproducible data on crumpled sheets.

We can predict the scaling relation between $M$ and $h$ if we assume a
small volume fraction and a uniform network of ridges between nearly flat
facets of characteristic size $X$ that depends on compression.  The energy
of each facet is proportional to $\kappa (X/\delta)^{1/3}$, where $\kappa$
is the bending modulus of the sheet and $\delta$ is its
thickness~\cite{1,2,3,4,5}.  Facets are oriented at random, so there is roughly
one facet per volume $X^3$.  Thus the energy contained in the volume $V$ is
$E \propto V/X^3 \kappa (X/\delta)^{1/3}$.    The space occupied by one
facet is large enough to contain a stack of $X/\delta$ facets so that $\phi
\equiv \delta /X$ is a volume fraction.  Macroscopically, $\phi = (\pi
\delta (L/2)^2)/(\pi h (D/2)^2)$, where $h$ is the height of the container,
$D$ is its diameter and $L$ is the diameter of the sheet.
$E$ can be re-expressed:
\begin{equation}
E = \kappa V/\delta ^3 (X/\delta)^{-8/3} = (\kappa /\delta^3) (\pi h
(D/2)^2) \phi^{8/3} \propto h^{-5/3}.
\label{eqn4}
\end{equation}
On the other hand, the energy $E$ supplied by the applied force $Mg$ is
given by  $E = \int_{\infty}^{h} M(h)g dh \approx Mgh$.  As noted
previously, we have assumed here that all the work goes into ridge energy.
We conclude that $Mgh \approx h^{-5/3}$, so that $h \propto M^{-3/8}$.
This predicted value of $\alpha = 0.375$ is to be compared with the
measured $\alpha = 0.54$.  A more careful estimate~\cite{15} yields $h
\approx 0.31 cm$ when $M = 0.1 kg$.  The predicted height, shown by the
dashed line of Fig.~\ref{fig3}(b), is a factor 1.5 to 3 lower than those
measured for the smallest masses, and approaches the measured heights for
the largest masses.

The computer simulations of Lobkovsky et al.~\cite{1} indicated that the
asymptotic scaling worked well when the volume fraction was no more that a
few tenths of a percent.   However, the experiments are well outside this
regime, where the volume fraction is large (over one half).  The size of
the observed facets is far from uniform, and this nonuniformity could alter
the stored energy. Further, we neglect friction which is expected to
strengthen the
crumpled sheet by a factor of order unity.  Plastic deformation is expected
to weaken it by at most a factor of order unity.  Considering these
differences between the experiment and the theoretical model, the agreement
between the two exponents is surprisingly good.

The data presented here suggests that the answer to our question in the
first paragraph, about how the ultimate size of a sheet depends on the
forces applied, has a complicated answer.  The $\log t$ dependence found in
the relaxation suggests that there is dissipation, either due
to friction or to plastic flow in regions of large curvature.  It is
unclear to what extent an analysis based only on conservative forces, which
does not consider such effects, can fully describe the energetics of
crumpling in real materials.  Despite that caveat, our data nevertheless do
support a scaling picture of the crumpling process: after a sheet has been
appropriately pre-treated, the height approaches an
asymptotic value as a power-law.

The results suggest several extensions such as studying how the exponent
$\alpha$ depends on size and thickness in different materials.  The most
pressing question is the origin of the logarithmic relaxation.  We do not
know if the behavior occurs because of plastic flow in the material or
because of frictional sliding at the contacts between sheets.  Neither
effect is included in the theoretic or simulation studies of crumpling.  In
order to determine the role of plastic flow in regions of large curvature,
one could measure the time-dependent relaxation in rubber sheets where such
plasticity would be minimized.

This research is supported by NSF  DMR-0089081 and MRSEC DMR-9808595.  We
thank A. Marshall for help in preparing this manuscript.

\end{document}